\documentstyle[psfig,aps,prl, multicol]{revtex}
\begin{document}
\title{Giant non-linearities  accompanying  
electromagnetically induced transparency}  
\author{Le-Man Kuang, Guang-Hong Chen,  and Yong-Shi Wu}
\address{Department of Physics, University of Utah, 
Salt Lake City, UT 84112, USA}
%\date{\today }
\maketitle
\begin{abstract}
We develop a fully quantum treatment of 
electromagnetically induced transparency 
(EIT) in a vapor of three-level $\Lambda$-type 
atoms. Both the probe and coupling lasers with
arbitrary intensities are quantized, and treated 
on the same footing. In addition to reproducing 
known results on ultraslow pulse propagation at
the lowest order in the ratio of their Rabi 
frequencies, our treatment uncovers that the 
atomic medium with EIT exhibits giant Kerr as 
well as higher order non-linearities. Enhancement 
of many orders of magnitude is predicted for 
higher-order refractive-index coefficients.
 
\end{abstract}

\begin{multicols}{2}

{\it Introduction} \hspace{0.10cm} 
The discovery of electromagnetically induced 
transparency (EIT) \cite{har} has led to the 
observation of new effects and development
of new techniques in quantum optics. Recent 
examples include ultraslow light pulse 
propagation \cite{hau,kas} and light storage 
\cite{liu,phi} in atomic vapor, and atomic 
ground state cooling \cite{mor}. It was proposed 
\cite{harr} that enhancement of efficiencies 
in nonlinear optical processes may be achieved 
by using EIT. A giant cross-Kerr non-linearity
in EIT was suggested by Schmidt and Imamo\v{g}lu 
\cite{sch}, and has been indirectly measured 
in the experiment \cite{hau}. We also note 
representative publications \cite{aga} 
utilizing EIT to study various optical 
non-linearities.

Essential to all these techniques is the 
phenomenon of EIT, in which coherent population 
trapping \cite{ari} is induced by atomic 
coherence and interference. As is well-known, 
a vapor of three-level $\Lambda$-type atoms, 
initially in the ground state, when irradiated 
by two laser beams (the coupling and probe 
laser), exhibits EIT under certain conditions.
 Conventionally, both the coupling and probe 
lasers were treated as classical, external 
fields. In such treatments, the occurrence
of EIT requires the coupling laser be much 
stronger than the probe laser. However, in 
the experiment \cite{hau} this condition was 
only marginally satisfied, which did not seem 
to have affected EIT. Moreover, with a quantum
treatment of the probe laser, Fleischhauer and 
Lukin \cite{fle} have recently been able to 
predict the possibility of coherently controlling 
the propagation of light pulses via dark-state 
polaritons, which are formed through quantum 
entanglement of atomic and probe-photon states. 
This possibility is realized in the latest 
light storage experiments \cite{liu,phi}. This 
lesson teaches us that the quantum description 
of laser is more fundamental than the classical 
one, having advantages in uncovering new effects 
involving quantum nature of photons.  

As a natural next step in this direction, in 
the present Letter we initiate a fully quantum 
treatment of EIT, namely we will deal with two 
quantized (coupling and probe) photon field 
modes interacting with three-level $\Lambda$-type 
atoms. In this treatment the probe and coupling 
lasers are more or less on the same footing. This 
will have advantages in studying non-linearities 
in EIT, which requires the capability to deal 
with higher orders in the ratio of probe to 
coupling laser strengths (or the ratio of Rabi 
frequencies). Indeed, we will see that in 
addition to reproducing the well-known results 
on ultraslow light pulse propagation in atomic 
vapor at the lowest order, our treatment uncovers 
giant higher-order optical non-linearities in 
the atomic medium with EIT. They give rise to 
dramatic enhancement of Kerr as well as  
higher-order refractive-index coefficients.
  
{\it The Model and Dressed States} 
\hspace{0.10cm} Let us consider a three-level 
atom, with energy levels $E_1<E_3<E_2$, 
interacting with two quantized laser fields, 
in the $\Lambda$-type configuration (see Fig. 1): 
The lower two levels $|1\rangle$ and $|3\rangle$ 
are coupled to the upper level $|2\rangle$. 
Initially the atom is in the ground state 
$|1\rangle$. First applied is the coupling 
laser of frequency $\omega_2 =(E_2-E_3)/\hbar$. 
%in resonance with the transition 
%$|2\rangle \to |3\rangle$. 
It prepares necessary atomic coherence 
allowing the later applied probe laser of 
frequency $\omega_1=(E_2-E_1)/\hbar-\Delta_1$ 
to pass through the atomic medium without 
much absorption, even if the detuning $\Delta_1$ 
is zero. This is the EIT \cite{har} underlying 
the experimentally observed ultraslow pulse 
propagation in atomic vapor\cite{hau,kas}. 
In the following  we will present a new treatment 
of EIT, in which both the probe and the coupling 
lasers are quantized.

\begin{figure}
\psfig{figure=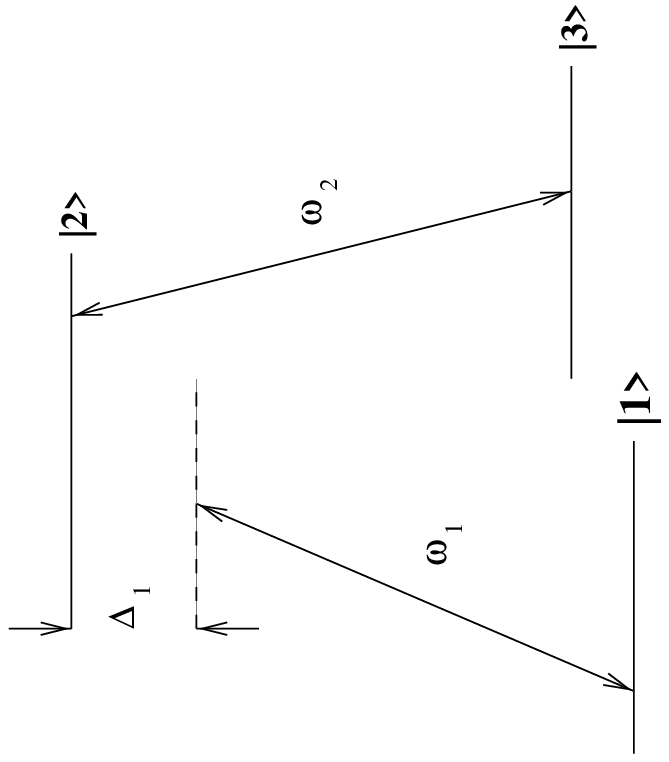,width=2.6in,height=1.50in,angle=-90}
\vskip 0.2cm
\center{Fig.~1.  Energy levels for a $\Lambda$-type atom.}
\end{figure}

With a unitary transformation 
$U(t)=\exp (-iH_0 t/\hbar)$, with $H_0= 
\sum_m E_m |m\rangle \langle m|- \hbar[\Delta_1 
(|2\rangle \langle 2| + |3\rangle \langle 3|)
+\omega_1\hat{a}^\dagger_1 \hat{a}_1 
+\omega_2\hat{a}^\dagger_2 \hat{a}_2]$, we
adopt an interaction picture, in which the 
Hamiltonian of the atom-field system in the 
rotating-wave approximation remains
time-independent:
\begin{eqnarray}
\label{e1}
\hat{H}&=&\hbar\Delta_1|2\rangle\langle 2| 
+ \hbar\Delta_1|3\rangle\langle3|
\nonumber \\
& &+ \hbar(g_1\hat{a}_1|2\rangle\langle 1|
+g_2\hat{a}_2|2\rangle\langle 3| + H.c. ),  
\end{eqnarray}
where $|m\rangle$ ($m=1,2,3$) are atomic 
states, $g$'s the coupling constants,  
and $\hat{a}_j$ and $\hat{a}^\dagger_j$
($j=1,2$) the annihilation and creation
operators of the probe and coupling laser 
modes. Assuming the detuning $\Delta_1$  
%coupling laser is resonant while the 
%probe is nearly resonant, 
is small, we can develop a perturbation 
theory in which we retain only terms 
linear in $\Delta_1$. The eigenstates of the
Hamiltonian (\ref{e1}), also called the dressed 
states, have the following form:
\begin{eqnarray}
\label{e2}
|\phi^{(i)}_{n_1, n_2}\rangle&=&
a_i |1,n_1, n_2\rangle
+b_i |2,n_1-1, n_2\rangle  
\nonumber \\
& &  +c_i |3,n_1-1, n_2+1\rangle,
\end{eqnarray}
where $i=\pm, 0$. The coefficients 
are given by 
\begin{eqnarray}
\label{e3}
a_{\pm} &=&\frac{\Omega_1}
{\sqrt2\Omega} \left [1\mp
\frac{\Omega^2_1 + 4\Omega^2_2}
{2\Omega^3}\Delta_1 \right ], 
\\
b_{\pm} &=&\pm\frac{1}{\sqrt2} 
\left [1\pm \frac{\Omega^2_1}{2\Omega^3}
\Delta_1\right ], \quad
c_{\pm} =\frac{\Omega_2}
{\sqrt2\Omega} \left [1\pm
\frac{3\Omega^2_1}{2\Omega^3}\Delta_1\right ],  
\nonumber
\end{eqnarray}
and
\begin{eqnarray}
\label{e4}
a_{0}=\frac{\Omega_2}{\Omega}, 
\;\;
b_{0} =\frac{2\Omega_1\Omega_2}
{\Omega^3}\Delta_1,\;\;
c_{0} =-\frac{\Omega_1}{\Omega}, 
\end{eqnarray}
with $\Omega_1=2g_1\sqrt{n_1}$, 
$\Omega_2=2g_2\sqrt{n_2+1}$ and
$\Omega=\sqrt{\Omega^2_1+\Omega^2_2}$.
The energy eigenvalues of the dressed states 
(\ref{e2}) are
\begin{eqnarray}
\label{e5}
E^{(\pm)}_{n_1, n_2}=\frac{\Omega^2_1
+ 2\Omega^2_2}{2\Omega^2}\Delta_1
\pm\frac{\Omega}{2}, \quad 
E^{(0)}_{n_1, n_2}=\frac{\Omega^2_1}
{\Omega^2}\Delta_1.
\end{eqnarray}

The complete set of dressed states for 
the system under consideration comprises 
the states $|\phi^{(i)}_{n_1,n_2}\rangle 
(i=\pm, 0)$ for each $n_1>0$, and  $n_2 \ge 0$ together 
with other two states $|1,0, n_2\rangle$ 
and $|3,n_1,0\rangle$ with zero eigenvalues. 
Knowing this set allows us to determine the 
time evolution of the atom-field system for 
any initial configuration. 

{\it Steady State and Susceptibility}
Assume that the atom is initially in the 
ground state, while the coupling and probe 
lasers in a coherent state $|\alpha, 
\beta\rangle$, with $\alpha$ and $\beta$ 
supposed to be real for simplicity. Namely 
the initial state of the atom-field system
is assumed to be
\begin{equation}
\label{e6}
|\Psi(0)\rangle =|1\rangle\otimes |\alpha,\beta\rangle.
\end{equation}
In the usual treatment with both lasers 
being classical external fields, one is 
concerned with a steady state of the 
atomic system. The counterpart of the 
steady state in our fully quantum treatment, 
according to the commonly used {\it adiabatic 
hypothesis} in quantum scattering theory
and quantum field theory\cite{gell}, is 
the state that evolves from the initial 
state (\ref{e6}) with the couplings $g_1$ and 
$g_2$ {\it adiabatically turned on}. 
Physically this is equivalent to having 
localized laser pulses before they enter 
the atomic vapor, with the pulse shape 
sufficiently smooth. In conformity to the 
Adiabatic Theorem, we need to identify 
a linear combination of the dressed
states (\ref{e2}) that tends to the initial 
state (\ref{e6}) if we take the limits $g_1,g_2 
\to 0$ (or $\Omega_1, \Omega_2 \to 0$). 
According to Eqs. (\ref{e3}-\ref{e5}), the ordering of 
the limits $\Omega_1 \to 0$ and $\Omega_2 
\to 0$ is important. Corresponding to 
the actual conditions in which EIT is 
observed, the correct ordering is first 
$\Omega_1 \to 0$ and then $\Omega_2 \to 0$. 
In our interaction picture, this procedure 
selects the $i=0$ state in Eq. (\ref{e2}). 
Transforming it back to the Schr\"{o}dinger 
picture, we identify the following state
as the state that evolves adiabatically 
from the initial state:
\begin{eqnarray}
 |\Psi(t)\rangle=\sum^{\infty}_{n_1,n_2=0}
e^{-(\alpha^2+\beta^2)/2}e^{-iEt}
\frac{\alpha^{n_1}\beta^{n_2}}{\sqrt{n_1!n_2!}}   
%\nonumber  \\
|\phi^{(0)}_{n_1,n_2}\rangle,
\label{e7}
\end{eqnarray} 
with $E= E_1+\omega_1n_1+\omega_2n_2
+E^{(0)}_{n_1,n_2}$. This state can be viewed 
as the counterpart of the usual "steady" state 
in our treatment. 
We see that the population of the upper level 
$|2\rangle$ is zero up to first order in detuning 
$\Delta_1$. This means that there is no 
absorption, implying the phenomenon of EIT. We 
note that for EIT to occur the usual treatment 
requires the coupling laser be much stronger 
than the probe one \cite{scu}, i.e. $\Omega_2 
\gg \Omega_1$; then most atoms are populated in 
the ground state. Our formalism does not require
this for EIT, so broader conditions are allowed: 
The coupling and probe lasers can be equally 
strong or even the probe laser is stronger. 
Then atoms can be in a more general superposition 
of two atomic states, which are entangled with 
the coupling and probe fields.  

From Eq. (\ref{e7}) one can form the total density 
operator of the atom-field system. After 
taking trace over the field states, one 
gets the atomic reduced density operator. 
Then the Fourier component of the optical 
coherence of the "steady" state at the 
frequency of the probe laser is found to be
\begin{equation}
\label{e8}
\rho^A_{21}(\omega_1)=a_0
(\bar{n}_{\alpha},\bar{n}_{\beta})b_0(\bar{n}_{\alpha},\bar{n}_{\beta}),
\end{equation} 
where $\bar{n}_{\alpha}=\alpha^2$, and  
$\bar{n}_{\beta}=\beta^2$ are, respectively, 
the mean photon numbers of the coupling 
and probe lasers in the coherent state 
$|\alpha, \beta\rangle$. In the derivation 
of Eq. (\ref{e8}), we have used the large-$n$ 
approximation, i.e., $\bar{n}_\alpha, 
\bar{n}_\beta \gg 1$, so that the photon
distributions are sharply peaked around 
their mean value. (For example, in a recent 
experiment of light storage \cite{liu}, 
$\bar{n}\sim 10^4$.)

The polarization of the atomic medium at 
the frequency of the probe laser is 
determined to be
\begin{equation}
\label{e9}
P(\omega_1)=\mu_{21}N\rho^A_{21}(\omega_1)
=\epsilon_0\chi(\omega_1)E_1(\omega_1), 
\end{equation}   
where $N$ is the number density of atoms, 
$\mu_{21}$ the transition dipole moment 
between states $|2\rangle$ and $|1\rangle$, 
$\epsilon_0$ the free space permittivity, 
$\chi(\omega_1)$ the susceptibility of the 
atomic medium, $E_1(\omega_1)$ the Fourier 
component of the mean electric field for 
the probe laser at frequency $\omega_1$. 
The value of $E_1(\omega_1)$ can be 
calculated from the "steady" state (\ref{e7}). 
In the large-$n$ approximation, 
$E_1(\omega_1)={\cal E}_1\alpha$, where ${\cal E}_i
=(\hbar\omega_i/2\epsilon_0V)^{1/2}$ ($i=1,2$) with  $V$ the 
quantized volume. Then the 
susceptibility is given by
\begin{equation}
\label{e10}
\chi(\omega_1)=\frac{4N|\mu_{12}|^2
\bar{\Omega}^2_2\Delta_1}{\hbar\epsilon_0
(\bar{\Omega}^2_1+\bar{\Omega}^2_2)^2},
\end{equation}  
where $\bar{\Omega}_1=\Omega_1(\bar{n}_{\alpha},
\bar{n}_{\beta})$ and $\bar{\Omega}_2=\Omega_2
(\bar{n}_{\alpha},\bar{n}_{\beta})$ are the 
Rabi frequencies of the coupling and probe 
lasers, respectively. This is the main result
of this paper, valid for arbitrary ratio of 
$\bar{\Omega}_1/ \bar{\Omega}_2$.

Eq. (\ref{e10}) exhibits the signature of EIT: 
the linear susceptibility vanishes 
at the resonance ($\Delta_1=0$). 
The derivative of $\chi (\omega_1)$
is related to the group velocity for
the probe laser pulse:
$v_g=c/[1+(\omega_1/2)(d\chi/d\omega_1)]$,
%evaluated at the probe frequency $\omega_1$ 
with $c$ speed of light in vacuum.
Thus we have 
\begin{equation}
\label{e11}
v_g=v^0_g \frac{\left (\bar{\Omega}^2_1
+\bar{\Omega}^2_2 \right )^2 }
{\bar{\Omega}^4_2},
\end{equation}
where $v^0_g=\hbar c\epsilon_0
\bar{\Omega}^2_2/(2\omega_1|\mu_{12}|^2N)$ 
is the usual expression for the group 
velocity \cite{hau}. This equation 
shows that in general the group velocity 
of the probe laser depends on the Rabi 
frequency of the coupling as well 
as the probe laser. The $\bar{\Omega}_1$ 
dependence of the group velocity is a new 
result of our treatment, closely related 
to the higher order nonlinear 
susceptibilities we will discuss below. 

To relate the present results to earlier 
studies of EIT-based dispersive properties 
\cite{harri}, we expand the susceptibility 
given in Eq. (\ref{e10}) in terms of powers 
of $ \bar{\Omega}_1/ \bar{\Omega}_2$. When 
$\bar{\Omega}_1 \ll \bar{\Omega}_2$, the 
first term of this expansion gives us the 
linear susceptibility
\begin{equation}
\label{e12}
\chi^{(1)}(\omega_1)=\frac{4|\mu_{12}|^2N
\Delta_1}{\hbar\epsilon_0\bar{\Omega}^2_2}.
\end{equation}
Thus, at the lowest order we recover results 
in Ref. \cite{harri}, when the decay rates 
of states are neglected at the resonant 
frequency of the probe laser. Similarly,
keeping only the lowest-order  term  in 
$\bar{\Omega}_1/\bar{\Omega}_2$, Eq. (\ref{e11}) 
will give the usual expression, $v_g^0$, for 
the group velocity \cite{hau}, which does not 
depend on $\bar{\Omega}_1$ and involves the 
contributions only from linear susceptibility.

The refractive index of the medium can be 
obtained from Eq. (\ref{e10}) by definition  
$n \equiv \sqrt{1+ \chi}$. Making use of Eqs. 
(\ref{e10}) and (\ref{e11}), we find the 
refractive index change near zero probe 
detuning to be 
\begin{equation}
\label{var}
\Delta n=\frac{\lambda_1}{2\pi}\frac{\Delta_1}{v_g},
\end{equation}
where $\lambda_1$ is the wavelength of probe laser.
It is worthwhile to note that this refractive 
index change involves the contributions of all 
orders of nonlinear susceptibilities due to the 
intensity dependence of the group velocity in 
(\ref{e11}).  For the slow light experiment 
\cite{hau}, with $v_g=17$ m/s and parameters 
$\Delta_1=1.3\times 10^6$ rad/s and 
$\lambda_1=589$ nm, we obtain from Eq.(\ref{var}) 
$\Delta n=7.2\times 10^{-3}$. This value agrees 
with the measured value in Ref. \cite{hau}.

{\it Giant Non-linearities} \hspace{0.10cm}
Nonlinearities play an important role not only 
in nonlinear optics but also in quantum optics. 
They may be used for generation of enhanced 
squeezing \cite{tom}, quantum computation and 
quantum teleportation \cite{vit}, and quantum 
nondemolition measurements \cite{bra}.
Since EIT takes place in the vicinity of 
atomic resonance, large nonlinearities are  
naturally expected. We note that in the 
conventional steady-state approach, it is 
difficult to obtain nonlinear susceptibilities
higher than $\chi^{(1)}$ and $\chi^{(3)}$. 
However, in our formalism we can get arbitrary 
higher-order nonlinear susceptibilities once 
for all. 

The higher-order nonlinear susceptibilities are
defined by
\begin{eqnarray} 
\chi(\omega_1)&=&\chi^{(1)}(\omega_1)
+\chi^{(3)}(\omega_1)|E(\omega_1)|^2 \nonumber \\
\label{e13}
& &+\chi^{(5)}(\omega_1)|E(\omega_1)|^4
+ \chi^{(7)}(\omega_1)|E(\omega_1)|^6 +\cdots
\end{eqnarray}
where $\chi^{(1)}(\omega_1)$ is the linear 
susceptibility given in Eq. (\ref{e12}), and
$\chi^{(k)}(\omega_1) (k \ge 3)$ represent 
the $k$th-order nonlinear susceptibility.
Assuming that the coupling laser is 
stronger than the probe, we expand the 
susceptibility (\ref{e10}) in powers of 
$\bar{\Omega}_1/\bar{\Omega}_2$. Rewriting 
this expansion in the form of Eq. (\ref{e13}), 
we find 
\begin{eqnarray} 
\chi^{(3)}(\omega_1)&=&-\frac{4\epsilon_0c}
{I_2}\chi^{(1)}(\omega_1),
 \hspace{0.3cm}
\chi^{(5)}(\omega_1)=-\frac{3\epsilon_0c}
{I_2}\chi^{(3)}(\omega_1),
\nonumber \\
\label{e14}
\chi^{(7)}(\omega_1)&=&-\frac{8\epsilon_0c}
{3I_2}\chi^{(5)}(\omega_1),
\end{eqnarray}
where $I_2=2\epsilon_0{\cal E}_2c\beta^2$ is 
the intensity of the incident coupling laser.

Normally nonlinear optical properties of
the medium are described by the nonlinear
refractive indices \cite{rat}:
\begin{eqnarray}
\label{e15} 
n&=&n_0+n_2|E_1|^2+n_4|E_1|^4 +n_6|E_1|^6 + \cdots
\end{eqnarray}
where the first nonlinear correction to the 
refractive index is the Kerr coefficient 
$n_2$, which is related to 
$\chi^{(3)}(\omega_1)$. 
$n_k (k\ge 4)$ are higher-order nonlinear 
refractive-index coefficients, related to 
higher-order nonlinear susceptibilities 
up to $\chi^{(k+1)}(\omega_1)$.

Making use of Eqs. (\ref{e13}) and (\ref{e14}), we obtain 
from Eq. (\ref{e15}) nonlinear refractive index 
coefficients:
\begin{eqnarray} 
n_2&=&-\frac{2\epsilon_0c}{I_2}\chi^{(1)}(\omega_1)
=-\frac{2\epsilon_0c\Delta_1}{\pi I_2} \frac{\lambda_1}{v^0_g}, 
\nonumber \\
\label{e16}
n_4&=&-\frac{3\epsilon_0cn_2}{I_2}, \hspace{0.5cm} 
n_6=-\frac{8\epsilon_0cn_4}{3I_2}.
\end{eqnarray}
%where $\lambda_1$ is the wave length of the probe laser.
To increase the value of nonlinear refractive-index 
coefficients one may either increase the atomic 
density or decrease the coupling laser intensity.

To demonstrate the magnitude of the giant 
nonlinearities derived above, we calculate 
the nonlinear refractive-index coefficients 
using the parameters ($I_2=40$ {\rm mW}/{\rm c$m^2$}, 
$\Delta_1=1.3\times 10^6$ rad/s, and 
$\lambda_1=589$ nm) in the ultraslow light 
experiment reported in Ref. \cite{hau}, in
which a light pulse speed $17$ m/s was 
observed. We estimate that under these
conditions, $n_2=-1.9\times 10^{-7}$ m$^2$/V$^2$, 
$n_4=3.8\times 10^{-12}$ m$^4$/V$^4$, and 
$n_6=-6.7\times 10^{-17}$ m$^6$/V$^6$. In 
terms of a common practical unit,   
$n_2=-0.36 {\rm cm}^2/{\rm W}$,
$n_4=13.0  {\rm cm}^4/{\rm W}^2$,
and $n_6=4.5 \times 10^2 {\rm cm}^6/{\rm W}^3$.
The value of the Kerr nonlinearity is of the 
same order of magnitude as that indirectly 
measured in Ref. \cite{hau}, almost $10^6$ times 
greater than that measured in cold Cs atoms 
\cite{hau,lam}, and $\sim 10^{12}$ times greater 
than that measured in other materials 
\cite{sal}. The fourth-order refractive-index 
coefficient $n_4$ is $\sim 10^{22}$ times 
greater than that measured in other 
materials \cite{sal}.

In optical fibers, the ratio between the 
second- and the fourth-order refractive-index 
coefficients is an essential parameter\cite{wri}  
to obtain stable  spatial solitary waves.  The lower 
the ratio $n_2/n_4$, the lower the required 
power for stable beam propagation. For the 
atomic medium with EIT, we have $n_2/n_4
=-I_2/(3\epsilon_0c)$. With the parameters in 
the experiment\cite{hau}, we estimate 
$|n_2/n_4|\sim 10^{-2} {\rm W}/{\rm cm}^2$.
This ratio is small compared to most other 
nonlinear media \cite{sal} by almost 11 orders 
of magnitude. From Eq. (\ref{e16}) we also 
obtain the ratio between the fourth- and the 
sixth-order refractive-index coefficients   
$n_4/n_6=-3I_2/8(\epsilon_0c)$, which is of 
the same order as $n_2/n_4$.

We note that when the ratio of $\bar{\Omega}_1/
\bar{\Omega}_2$ is close to unity, one had better
directly deal with  the original formula (10), 
rather than using its power series expansion. 

{\it Conclusions} \hspace{0.10cm} We have
developed a fully quantum treatment of EIT
in atomic medium. Both the probe and coupling
lasers are quantized and treated on the same 
footing. This allows us to deal with the cases
in which the ratio of the probe-to-coupling 
Rabi frequencies is not necessarily small. 
At the lowest order in this ratio, we are 
able to reproduce the known results for slow
pulse propagation. At higher orders we have
uncovered that accompanying the EIT phenomenon,
atomic medium possesses giant optical 
non-linearities, which can give rise to dramatic
enhancement of the Kerr as well as higher order
refractive-index coefficients. In other words,
the atomic medium with EIT is really a very 
unusual optical medium; to the list of its 
astonishing optical properties, we now add
another remarkable one: giant optical
non-linearities. It would be very interesting 
to observe  these giant non-linearities and 
explore their potential applications.

Contrary to the usual treatment of EIT, 
our present treatment is essentially a 
time-independent approach, because both 
probe and coupling lasers are included 
as part of the dynamical system. On the 
other hand, an important limitation is 
that we have ignored the decay rates of 
various levels. How to incorporate them 
in a fully quantum treatment is a 
challenging issue. 

This work was supported in part by the US 
NSF under grant PHY-9970701, and a Seed 
Grant from the University of Utah. L.M.K. 
also acknowledges support from the Climbing 
Project, NSF and EDEYTF of China, and  
Hunan Province STF.

\end{multicols}

\end{document}